\documentclass[12pt]{iopart}

\usepackage{iopams}
\begin{document}

\title[Noether symmetry in quintom cosmology]{Revisiting Noether gauge symmetry approach in quintom cosmology}

\author{Sajid Ali}

\address{Dept. of Basic Sciences, School of Electrical Engineering and Computer Science, National University of Sciences and Technology, Campus H-12,
Islamabad 44000, Pakistan}
\ead{sajid$\_$ali@mail.com}
\vspace{10pt}
\begin{indented}
\item[]February 2014
\end{indented}

\begin{abstract}
The Noether gauge symmetry approach is revisited to study various quintom scenarios (those that arise by the presence of two dynamical scalar fields) to comprehend the role of dark energy in our universe. For such models, we obtain smooth parameterizations of the equation of state of dark energy across the boundary of cosmological constant $w_{\Lambda}=-1$. This study gives rise to two new cases of the potential $V(\phi, \sigma)$, due to a quintom field in which nonlinear coupling of the scalar fields arise. Besides we report that a few cases of Noether gauge symmetries and their invariants in [Adnan Aslam, et. al., Astrophys Space Sci (2013), 348:533-540] are incorrect. Consequently, the given cosmological model in their paper is not a feasible quintom model.
\end{abstract}

%
%
%
%
%

\section{Introduction}

Noether symmetries play an essential role in finding conservation laws (or invariants) of any physical system using Noether's first theorem \cite{noe18,tav71}. For example, invariance of a Lagrangian of a physical system (in which force is due to the potential in the field, i.e., conservative systems) under time translation and rotation correspond to conservation of energy and angular momentum in favorable boundary conditions. Noether symmetry approach provides an elegant way  to find conserved quantities for a given Lagrangian. Noether symmetries for a given Lagrangian generate possible forms of the unknown function of the field variables which help in reducing the dynamics and to find exact solutions. In \cite{cap08}, the authors have used Noether symmetry approach to get possible cosmological solutions in $f(R)-$gravity. On the other hand, Noether gauge symmetry approach has an edge over standard Noether symmetry approach as it unveils more conserved quantities than from the other approach. This is clear from the observation that certain conserved quantities do not correspond to invariance of Lagrangian under continuous groups of transformations but under gauge transformations. For example, conservation of charge in quantum field theory correspond to a gauge invariance of the given Lagrangian.

An important problem in modern cosmology is to comprehend the role of dark energy that is now supported by observational evidence that at present the universe is going through cosmic acceleration at a fast pace. There are three important evidences to support above argument based on the experimental study of (a) supernovae Ia (SNIa) \cite{ri98,per99} (b) cosmic microwave background radiation along with large-scale structure surveys (CMB $\&$ LSS) \cite{kom09} and finally (c) the age of our universe calculated by incorporating dark energy \cite{ca10}. Einstein's general relativity at large cosmological scales requires that our universe must carry two special features, i.e., it is both homogeneous and isotropic. Therefore, our universe has a mysterious form of matter whose role is opposite of gravity such that its pressure is negative that accounts for this acceleration. Observations confirm the energy density of dark energy occupies 70$\%$ of our universe. It is also clear that the role of this mysterious matter at an earlier cosmological epoch is opposite to the present epochs because otherwise it may not have been possible to see present large-scale structures of our universe.

Equation of state $w_{DE} = p_{DE}/\rho_{DE}$, where $p_{DE}$ and $\rho_{DE}$ is the pressure and mass density, characterizes dark energy. For example, the case of a nonzero and positive cosmological constant boundary corresponds to $w_{\Lambda}=-1$, in which case $\rho_{\Lambda}$ is independent of the scale factor $a(t)$. A quintessence field is a dynamical field for which evolution of its equation of state evolves in the range $w_{Q}\geq -1$. Similarly for a phantom field, $w_{P}\leq -1$. There are evidences that the cosmological boundary is crossed \cite{hut05} therefore one would be interested in knowing a dynamical field for which equation of state is able to evolve the cosmological constant boundary $w_{\Lambda}=-1$. On the other hand, there exists a no-go theorem that forbids the equation of state of a single scalar field to cross over the cosmological constant boundary \cite{xi08}. One possible solution to this problem is to introduce a combination of two dynamical scalar fields, i.e., a canonical field $\phi$ and a phantom field $\sigma$. Such phenomenological models are known as quintom models which give rise to quintom cosmology in which the dynamics of dark energy is understood by using parameterizations of its equation of state \cite{fen05,guo05,joy14}.

In this paper, our purpose is twofold. First we point out all the errors in \cite{ad13} in the next section. Secondly we discuss some more cases in the third section that arise from Noether gauge symmetry approach in which the dynamical evolution of combination of the two fields (in the form of nonlinear sum or difference) yield interesting quintom models. Later we discuss useful insights in quintom cosmology obtained from those new solutions.

\section{Basic Quintom Model}
For a simple double field ($\phi,\sigma$) quintom model with an arbitrary potential $V(\phi,\sigma)$ \cite{fen05,guo05}, the action integral has the form
\begin{equation}
S= \frac{1}{2} \int d^{4}x  \sqrt{-g} \left [ R + \phi_{;\mu} \phi^{;\mu} - \sigma_{;\mu} \sigma^{;\mu} -2 V(\phi, \sigma) \right ], \label{eq1}
\end{equation}
where $g$ denotes the determinant of the Riemannian metric of spatially flat FRW metric $g_{\mu\nu} = \mbox{diag} \left ( 1, -a(t)^2 \sum_{3} \right)$, and where metric of the three dimensional Euclidean space is $\sum_{3}$. Further, $a(t)$ is the usual scale factor describing the expansion of our universe. The field variables $\phi(t)$ and $\sigma(t)$ correspond to dynamical canonical and phantom fields, respectively \cite{ca10}. It is important to include all physical equations in our analysis so the equations of motions are given by
\begin{eqnarray}
2\frac{\ddot{a}}{a} + \left ( \frac{\dot{a}}{a} \right )^2 = -p, \nonumber \\
\ddot{\phi} +3H \dot{\phi} + \frac{\mbox{d}V}{\mbox{d}\phi} = 0 , \nonumber \\
\ddot{\sigma} +3H \dot{\sigma} - \frac{\mbox{d}V}{\mbox{d}\sigma} = 0 . \label{eom}
\end{eqnarray}
The effective energy density and effective pressure are
\begin{eqnarray}
\rho = \frac{1}{2} \dot{\phi}^2 - \frac{1}{2} \dot{\sigma}^2 +V(\phi, \sigma), \\
p = \frac{1}{2} \dot{\phi}^2 - \frac{1}{2} \dot{\sigma}^2 - V(\phi, \sigma).
\end{eqnarray}
Thus we obtain the equation of state
\begin{equation}
w= \frac{\dot{\phi}^2 - \dot{\sigma}^2 - 2V}{\dot{\phi}^2 - \dot{\sigma}^2 + 2V}~. \label{eos}	
\end{equation}
The necessary condition for any feasible quintom model requires that close to the cosmological constant boundary $w=-1$, we must have $w^{\prime} |_{w=-1} \neq 0$, in which case the dynamical evolution of both fields generates a quintom scenario such that the boundary is crossed smoothly. Due to its significance we obtain a general form of $w^{\prime}$, using above definitions of effective pressure $P$ and effective energy density $\rho$, along with equations of motions (\ref{eom}). By differentiating (\ref{eos}), with respect to $t$, we get
\begin{equation}
w^{\prime} = \frac{ 8(\dot{\phi}\ddot{\phi} - \dot{\sigma}\ddot{\sigma}) V - 4(\dot{\phi}^2-\dot{\sigma}^2  )(\dot{\sigma}V_{\sigma} + \dot{\phi}V_{\phi}) }{(\dot{\phi}^2 - \dot{\sigma}^2 +2V)^2} ~,
\end{equation}
which upon using equations of motions (\ref{eom}), becomes
\begin{equation}
w^{\prime} = \frac{-24VH(\dot{\phi}^2 - \dot{\sigma}^2) - 4(\dot{\phi}^2 - \dot{\sigma}^2+2V)(\dot{\phi}V_{\phi} + \dot{\sigma}V_{\sigma})}{(\dot{\phi}^2 - \dot{\sigma}^2 +2V)^2} ~. \label{dw}
\end{equation}
The above equation can be used to keep a consistency check on the cases especially when closed form of the solutions of equations of motions can not be obtained. We now discuss all possible cases in which equations of motions can be integrated explicitly with the help of Noether gauge conditions.

In \cite{ad13}, the authors have used the Noether gauge symmetry approach to find Noether symmetries of the Lagrangian $L(a,\dot{a},\phi, \dot{\phi}, \sigma, \dot{\sigma})$ of (\ref{eq1}), given by
\begin{equation}
L = -3a \dot{a}^2 + a^{3} \left( \frac{1}{2}\dot{\phi}^2 -
\frac{1}{2} \dot{\sigma}^{2} - V(\phi,\sigma) \right) . \label{lag}
\end{equation}
The main objective of the paper was to obtain possible forms of the potential $V(\phi,\sigma),$ of a quintom field which are allowed by the presence of Noether symmetries. The Noether gauge symmetries are defined in terms of the vector fields
\begin{equation}
X = \mathcal{T} \frac{\partial}{\partial t} + \alpha \frac{\partial}{\partial a} + \beta \frac{\partial}{\partial \phi}+\gamma \frac{\partial}{\partial \sigma} ~, \label{sym}
\end{equation}
provided they satisfy the Noether gauge symmetry conditions
\begin{equation}
X^{(1)} L + L(D_{t} \mathcal{T}) = D_{t} G, \label{cond}
\end{equation}
where $G(t,a,\phi,\sigma),$ is an unknown function called the gauge function of the corresponding symmetry. Using Noether theorem the conserved quantity can be calculated from the formula
\begin{equation}
I = \mathcal{T}	L + (\eta^{\mu} - \dot{x}^{\mu}\mathcal{T}) \frac{\partial L}{\partial \dot{x}^{\mu}} - G , \label{inv}
\end{equation}
where $x^{\mu}=(a,\phi, \sigma)$ and $\eta^{\mu} = (\mathcal{T},\alpha,\beta,\sigma)$ denote the variables of Lagrangian (\ref{lag}) and coefficients of the Noether symmetry (\ref{sym}), respectively. Note that $D_{t}I=0$, upon using the equations of motions, where
\begin{equation}
D_{t} = \frac{\partial}{\partial t} + \dot{a}\frac{\partial}{\partial a} +\dot{\phi}\frac{\partial}{\partial \phi} +
\dot{\sigma}\frac{\partial}{\partial \sigma}~.
\end{equation}
We first obtain the determining equations and give a list of the cases of Noether symmetries from \cite{ad13}. We check the validity of each of their cases and show that in some cases the equations of motions are not satisfied. 

\subsection{List of Cases}
The equation (\ref{cond}) yields a system of linear partial differential equations (PDEs) to be solved simultaneously. This is system (10) in \cite{ad13}, which we include here for completeness
\begin{eqnarray}
\mathcal{T}_{a} =0,\mathcal{T}_{\phi} =0, \mathcal{T}_{\sigma} =0, \label{sys1}\\
\beta_{\sigma} - \gamma_{\phi} =0, \label{sys2}\\
6\alpha_{\sigma} -a^2 \gamma_{a} =0, \label{sys3}\\
-6\alpha_{\phi} + a^2 \beta_{a} =0, \label{sys4}\\
a^3\gamma_{t} +G_{\sigma} =0, \label{sys5}\\
a^{3}	\beta_{t} -G_{\phi} =0, \label{sys6}\\
6a \alpha_{t} +G_{a} =0, \label{sys7}\\
\alpha + 2a\alpha_{a} -a\mathcal{T}_{t} =0, \label{sys8}\\
3\alpha + 2a\beta_{\phi}	-a\mathcal{T}_{t} =0, \label{sys9}\\
3\alpha + 2a\gamma_{\sigma}	-a\mathcal{T}_{t} =0, \label{sys10}\\
3a^{2}V\alpha +a^3\beta V_{\phi} + a^3 \gamma V_{\sigma} + a^{3} V \mathcal{T}_{t} +G_{t} =0. \label{sys11}
\end{eqnarray}
The above system was solved for different cases in \cite{ad13}. We show that the cases (4) and (5) given in their paper, are incorrect while case (2) is incomplete. Further the given potential (20) in Section 5 (Cosmography) \cite{ad13} fails to satisfy the equations of motions. Below we list these cases in detail.

1. In case (4), they have obtained a Noether symmetry
\begin{equation}
X= \frac{\partial}{\partial t} + \frac{1}{a^3}\frac{\partial}{\partial \phi}
\end{equation}
corresponding to the gauge function and potential
\begin{eqnarray}
&G = c_{1} t + c_{2} ,\\
&V(\phi,\sigma) = -c_{1} \phi +F(\sigma).
\end{eqnarray}
From equation (\ref{sym}), we find that for this symmetry $\mathcal{T}=1, \alpha =0, \beta = 1/a^3 , \gamma =0$, which clearly do not satisfy equation (\ref{sys4}) of the determining equations of Noether symmetries because $\alpha_{\phi} =0$ but $\beta_{a}= -3/a^4$, which can not be zero and leads to a contradiction.

2. Similarly the two Noether symmetries in case (5) are
\begin{eqnarray}
&X_{1}= \frac{\partial}{\partial t}~, \\
&X_{2}= \left (c_{1} \frac{t}{a^3}+F(a) \right)\frac{\partial}{\partial \phi}~,
\end{eqnarray}
corresponding to the gauge function and potential
\begin{eqnarray}
&G = c_{1} \phi + c_{2} ,\\
&V(\phi,\sigma) = F(\sigma).
\end{eqnarray}
If we choose $G$ of the above form then equation (\ref{sys6}), implies $a^{3}\beta_{t} - c_{1}=0$, which is not identically satisfied by the first symmetry $X_{1}$, for which $\beta =0$, therefore $c_{1}=0$ which makes the gauge independent of $\phi$. Similarly, for $X_{2},$ we have $\beta = (c_{1}t/a^3+F(a))$ and $\alpha=0,$ whereas equation (\ref{sys4}) implies that $\beta$ can no longer be a function of parameter $a$. Therefore both cases are incorrect and the resulting operators are not Noether gauge symmetries of the corresponding Lagrangian.

3. The case (2), is not complete as it contains two arbitrary constants whose ranges are not specified. Furthermore possible forms of the arbitrary function are not characterized
\begin{equation}
V= V(\phi, \sigma) = 	F \left ( \frac{1}{2}c_{1}(\sigma^2 - \phi^2) +c_{2} \sigma -c_{3} \phi \right ).
\end{equation}
For example, if we choose $c_{1} =0, c_{2} =1, c_{3} = \pm 1$, and the function $F$ takes the form given below then the Lagrangian has an extra Noether symmetry.

4. In Section 4 (Exact solutions), authors have used case (4) with $c_{1}=0$, to obtain an invariant of the Noether symmetry. They have also given Euler-Lagrange equations. This choice of the value of constant makes the gauge function $G=c_{2}$, and potential function as $V=F(\sigma)$. As we have proved that the given Noether symmetry for this case in \cite{ad13}, is not a Noether symmetry therefore the resulting invariant (equation (18) of their paper), namely
\begin{eqnarray}
I = 3a \dot{a}^2 - \frac{1}{2}	a^3 \dot{\phi}^2 + \frac{1}{2} a^{3} \dot{\sigma}^2 - a^3 F(\sigma) + \dot{\phi},
\end{eqnarray}
is not an invariant because $D_{t}I,$ on using the Euler-Lagrange equations
\begin{eqnarray}
& 4a\ddot{a} + a^{2} ( \dot{\phi}^2 - \dot{\sigma}^2) +2a^2 F(\sigma) -2\dot{a}^2 =0,  	 \\
& \ddot{\phi} -\frac{3\dot{a}}{a} 	\dot{\phi}  =0, \\
& \ddot{\sigma} - \frac{3\dot{a}}{a} \dot{\sigma} - \frac{d F(\sigma)}{d\sigma} =0,
\end{eqnarray}
is a non-zero factor given by
\begin{equation}
D_{t}I = -\frac{3\dot{a}\dot{\phi}}{a}~.
\end{equation}
This is zero provided $\phi=\mbox{constant.}$ They have obtained two exact solutions for this value of $\phi$, with further choices on
$F(\sigma) =2$ and $F(\sigma)=0$, both of which become sub cases of case (6). The case $F(\sigma)=c\neq 0$, gives four Noether symmetries whereas $F(\sigma) =0,$ gives five Noether symmetries
\begin{eqnarray}
&X_{1}= \frac{\partial}{\partial t}~, \quad X_{2}= 3t\frac{\partial}{\partial t}+a\frac{\partial}{\partial a}~, \\
&X_{3}= \sigma \frac{\partial}{\partial \phi} + \phi \frac{\partial}{\partial \sigma}~,\quad X_{4}=  \frac{\partial}{\partial \phi} , \quad X_{5}= \frac{\partial}{\partial \sigma}~.
\end{eqnarray}
It is a case when the potential $V(\phi,\sigma)$, is zero. For completeness we include their invariants
\begin{eqnarray}
I_{1} &= \frac{a}{2}\left ( 6\dot{a}^2 -a^2 ( \dot{\phi}^2 - \dot{\sigma}^2) \right), \\
I_{2} &= \frac{3a}{2}\left ( 6t\dot{a}^2 - 4a \dot{a} - ta^2 ( \dot{\phi}^2 - \dot{\sigma}^2) \right), \\
I_{3} &= a^3 ( \phi \dot{\sigma} - \sigma \dot{\phi} ) , \\
I_{4} &= a^3 \dot{\phi}, \\
I_{5} &= -a^3 \dot{\sigma},
\end{eqnarray}
and discuss possible exact solutions as well as their cosmological implications in Section 4. Similarly for $F \neq 0$, there are four Noether symmetries $X_{1},X_{3},X_{4},X_{5}$ and correspondingly there are four invariants
\begin{eqnarray}
I_{1} &= \frac{a}{2}\left ( 6\dot{a}^2 -a^2 ( \dot{\phi}^2 - \dot{\sigma}^2 + 2C) \right), \\
I_{3} &= a^3 ( \phi \dot{\sigma} - \sigma \dot{\phi} ) , \\
I_{4} &= a^3 \dot{\phi}, \\
I_{5} &= -a^3 \dot{\sigma}.
\end{eqnarray}

5. In Section 5 (Cosmography), the only potential taken in their paper was expected to reveal some new results but it contains serious flaws. The potential was confined to have the form
\begin{equation}
V(\phi,\sigma) = \frac{1}{2} (\sigma - \phi ) (\sigma + \phi + 1),
\end{equation}
with some specific choice of constants. The above function has the derivatives
\begin{equation}
\frac{\mbox{d}V}{\mbox{d}\phi} = -\frac{1}{2}(1+2 \phi) , \quad \frac{\mbox{d}V}{\mbox{d}\sigma} = \frac{1}{2}(1+2 \sigma),
\end{equation}
therefore, in their paper, equations of motions (21) and (22) which come from equations (3) and (4) are wrong. Consequently, equations (24) and (25) are also incorrect. Hence the subsequent analysis is totally unnecessary and contains fallacies in all diagrams and cosmological implications in Section 6 (Stability).

\section{Additional Cases}
The system of determining equations ($\ref{sys1})-(\ref{sys11}$), is difficult to solve in general. We have used Computer
Algebra System (CAS) Maple$-$17, to carry out case-splitting with
the help of an important algorithm `rifsimp' that is essentially an
extension of the Gaussian elimination and Groebner basis algorithms
that is used to simplify overdetermined systems of polynomially
nonlinear PDEs or ODEs and inequalities and bring them into a useful
form. It turns out that the system is inconsistent with the choice of a time dependent gauge function.  We have also verified all the cases in the previous section. Furthermore we obtain some intriguing forms of the potential function with the use of CAS that can shed light on the physical insights of canonical and phantom fields and their behaviour at different epochs of our universe. To the author's knowledge these results of dynamical invariants have not been reported before. Below we list these new cases.

1. We now find a solution of system ($\ref{sys1})-(\ref{sys11}$), in which a time varying invariant is obtained for a new quintom model. It is the case of a constant gauge function and potential $V(\phi,\sigma)$ with a nonlinear power of sum of the canonical field ($\phi$) and phantom field ($\sigma$), i.e., $\phi + \sigma$. Since we obtain a similar result in the case of difference of both fields ($\phi-\sigma$) with a change of plus/minus sign, therefore we include both in a unified form given below
\begin{eqnarray}
&
G = c,  \label{gau}\\
&
V(\phi, \sigma) = (\phi \pm \sigma) ^{n}, \quad n\neq 0, \label{newg}
\end{eqnarray}
where $n$ is a non-zero integer. It would be interesting to see the impact of above potential due to the presence of both canonical and phantom fields on the density and pressure of the universe \cite{ca10}. The resulting model is such that the sum (or difference) of contribution of canonical and phantom fields on the potential is increasing for $n>0$, and decreasing for $n<0$. Since the fields are time-varying in general therefore it could be that their sum (or difference) is increasing for certain periods of time and decreasing for other periods of time. The value $n=0$, corresponds to the case (6) in \cite{ad13}, that contains a typo, namely the Noether symmetry $X_{4}=F(a)\partial_{\phi},$ is incorrect as it contain $F(a)$ which is not possible. It is $X_{4} = \partial_{\phi}$.
We now obtain three Noether symmetries of the Lagrangian (\ref{lag}) for the above forms (\ref{gau}) $\&$ (\ref{newg}) of the gauge and potential functions which are
\begin{eqnarray}
&X_{1}= \frac{\partial}{\partial t}~, \quad X_{2}=  \frac{\partial}{\partial \phi} \mp \frac{\partial}{\partial \sigma}~, \nonumber \\
&X_{3}= t \frac{\partial}{\partial t} + \frac{a}{3} \frac{\partial}{\partial a}  \mp
\frac{2}{n} \left ( \sigma \frac{\partial}{\partial  \phi}  + \phi \frac{\partial}{\partial \sigma} \right ),
\label{nsym1}
\end{eqnarray}
where these satisfy the commutator relations
\begin{eqnarray}
&\big [ X_{1}, X_{2} \big ] = 0 ,  \big[ X_{1}, X_{3}\big] = X_{1} , \big[ X_{2}, X_{3}\big] = \pm \frac{2}{n} X_{2}.
\end{eqnarray}
The Noether symmetry $X_{1}$ corresponds to translations in time whereas $X_{2}$ is the difference (and respectively sum) of the translations in the field variables $\phi$ and $\sigma$. Interestingly, $X_{3}$ is a non-trivial Noether symmetry which includes a mixture of both scaling (in $t$ and $a$) and a hyperbolic rotation (in $\phi$ and $\sigma$).

Noether theorem ensures that there exist three invariants (conserved quantities) corresponding to three Noether symmetries for the Lagrangian (\ref{lag})
\begin{equation}
L = -3a \dot{a}^2 + a^{3} \left( \frac{1}{2}\dot{\phi}^2 -
\frac{1}{2} \dot{\sigma}^{2} - (\phi \pm \sigma )^n \right) . \label{lag1}
\end{equation}
From equation (\ref{inv}), the three invariants relative to each $X_{1}, X_{2}$ and $X_{3}$ of the above Lagrangian are
\begin{eqnarray}
&I_{1} =  3a\dot{a}^2 + a^3 \left (\frac{ \dot{\sigma}^2-\dot{\phi}^2 }{2} - (\phi \pm\sigma)^n \right), \\
&I_{2} =  a^{3} \left( \dot{\phi} \pm \dot{\sigma} \right),\\
&I_{3} =  2a^2 \dot{a} - 3ta\dot{a}^2 + \nonumber
\\
&a^{3} \left( t\left((\phi\pm\sigma)^n + \frac{\dot{\phi}^2 -\dot{\sigma}^2}{2} \right) \mp \frac{2}{n}\left (\phi \dot{\sigma} - \sigma \dot{\phi}\right ) \right).
\end{eqnarray}
The first two invariants could be guessed from the form of the Lagrangian (\ref{lag1}). However the third invariant is a  non-trivial dynamical invariant and represents a quantity which changes over time but remains conserved during evolution. Therefore both $I_{1}$ and $I_{2}$ are first integrals whereas $I_{3}$ is a constant of motion.

2. We also obtain another case in which the system has a solution with the gauge and potential functions of the form
\begin{eqnarray}
&
G = c, \label{new2} \\
&
V(\phi, \sigma) = \frac{(\phi \mp \sigma) ^{n}}{(\phi \pm \sigma) ^{m}}, \quad n\neq 0,m\neq 0, \quad n\neq -m \label{new3}
\end{eqnarray}
where $n$ and $m$ are two positive integers, and we again write down two separate cases in a unified form. The behavior of fields is more interesting in this case, for example if $n=m,$ then $V$ is the ratio of the difference to the sum of the fields (or in the other case it is the ratio of the sum to the difference of the fields). For $n\neq m$, we obtain two distinct cases. If $n>m,$ then the impact on potential of difference of the fields is larger than their sum (or impact due to sum of the fields is larger than their difference in the other case). On the other hand if $n<m$, then the impact of difference of the fields is smaller than their sum (or impact due to sum of the fields is smaller than their difference in the other case).
The two Noether symmetries of the Lagrangian (\ref{lag}) for the above forms (\ref{new2}) $\&$ (\ref{new3}) of the gauge and potential functions are
\begin{eqnarray}
&X_{1}= \frac{\partial}{\partial t}, X_{2}= t \frac{\partial}{\partial t} + \frac{a}{3} \frac{\partial}{\partial a}  \pm
\frac{2}{n+m} \left ( \sigma \frac{\partial}{\partial  \phi}  + \phi \frac{\partial}{\partial \sigma} \right ), \label{nsym2}
\end{eqnarray}
where the algebra is closed
\begin{eqnarray}
&\big [ X_{1}, X_{2} \big ] = X_{1}.
\end{eqnarray}
The two invariants corresponding to two Noether symmetries for the Lagrangian (\ref{lag})
\begin{equation}
L = -3a \dot{a}^2 + a^{3} \left( \frac{1}{2}\dot{\phi}^2 -
\frac{1}{2} \dot{\sigma}^{2} - \frac{(\phi \mp \sigma )^n}{(\phi \pm \sigma )^m} \right) , \label{lag2}
\end{equation}
have the form
\begin{eqnarray}
&I_{1} =  3a\dot{a}^2 + a^3 \left (\frac{ \dot{\sigma}^2-\dot{\phi}^2 }{2} -  \frac{(\phi \mp \sigma )^n}{(\phi \pm \sigma )^m}\right), \\
&I_{2} =  2a^2 \dot{a} - 3ta\dot{a}^2 + \nonumber
\\
&a^{3} \left( t\left(\frac{(\phi \mp \sigma) ^{n}}{(\phi \pm \sigma) ^{m}} + \frac{\dot{\phi}^2 -\dot{\sigma}^2}{2} \right)
\pm \frac{2}{n+m}\left (\phi \dot{\sigma} - \sigma \dot{\phi}\right ) \right). \label{inm}
\end{eqnarray}
In this case $I_{1}$ is the first integral whereas $I_{2}$ is a constant of motion.
\section{Quintom Solutions}

\textbf{Case 1.} ($V(\phi,\sigma)=0$)\newline
This is the case in the absence of a potential field, i.e., $V(\phi,\sigma)=0,$ in which we obtain five invariants corresponding to five Noether symmetries possessed by the Lagrangian (\ref{lag}). We include this case to verify previous cases as well to comprehend the role of invariants $I_{i},$ and constant gauges that result the evolution of both fields. There are five invariants for the case of constant gauge function, therefore we write
$I_{i}= c_{i},$ where $i=1,...,5$. We use small letters to denote constancy of invariants and integration constants are represented by capital letters. Note that these constants may carry any constant value nevertheless non-trivial solutions of equations of motions  depend if these constants are zero or not. For example, the invariant $I_{4}=c_{4}$, gives $\phi = \mbox{constant},$ if $c_{4}=0$ and $\dot{\phi} = c_{4}/a^3$ otherwise. Their values are specified by certain physical constraints. To obtain all possible cases exhaustively we have carried out case splitting using computer algebra system and found following solutions. 	  \newline
\textbf{1a. (i)} ($c_{i}\neq 0$ for all $i$)
\begin{eqnarray}
a(t) &= \frac{1}{2} \left (6c_{1}t^2 - 4c_{2}t + C_{1} \right ) ^{1/3}, \\
\sigma(t) &= - \frac{4c_{1}c_{5}}{ \sqrt{6(c_{5}^2 -c_{4}^2) }}\arctan{\left( \frac{3c_{1}t - c_{2}}{\sqrt{6(c_{5}^2 -c_{4}^2) }}\right)} + C_{2} ,\\
\phi(t) & = \frac{4c_{4}c_{1}}{\sqrt{6(c_{5}^2 -c_{4}^2) }}\arctan{\left( \frac{3c_{1}t - c_{2}}{\sqrt{6(c_{5}^2 -c_{4}^2) }}\right)} + \frac{c_{3}-c_{4}C_{2}}{c_{5}} , \\
&\mbox{where} \quad C_{1} = \frac{c_{2}^2 + 6 (c_{5}^2 -c_{4}^2)}{12 c_{1}}, \quad |c_{5}|>|c_{4}|
\end{eqnarray}
In the absence of potential ($V$) the dynamical $w$ reduces to a constant value
\begin{equation}
w= \frac{\dot{\phi}^2 - \dot{\sigma}^2}{\dot{\phi}^2 - \dot{\sigma}^2}=1~.
\end{equation}
Thus the simplest case yields $p = \rho$, which corresponds to a radiation model and the evolution of both fields is determined by the forms above. Subsequently in all sub-cases of case 1, the equation of state possesses the same form. \newline
\textbf{1a. (i)} ($c_{i}\neq 0$ for all $i$)
\begin{eqnarray}
a(t) &= \frac{1}{2} \left (6c_{1}t^2 - 4c_{2}t + C_{1} \right ) ^{1/3}, \\
\sigma(t) &= - \frac{4c_{1}c_{5}}{ \sqrt{6(c_{5}^2 -c_{4}^2) }}\arctan{\left( \frac{3c_{1}t - c_{2}}{\sqrt{6(c_{5}^2 -c_{4}^2) }}\right)} + C_{2} ,\\
\phi(t) & = \frac{4c_{4}c_{1}}{\sqrt{6(c_{5}^2 -c_{4}^2) }}\arctan{\left( \frac{3c_{1}t - c_{2}}{\sqrt{6(c_{5}^2 -c_{4}^2) }}\right)} + \frac{c_{3}-c_{4}C_{2}}{c_{5}} , \\
&\mbox{where} \quad C_{1} = \frac{c_{2}^2 + 6 (c_{5}^2 -c_{4}^2)}{12 c_{1}}, \quad |c_{5}|>|c_{4}|
\end{eqnarray}
In the absence of potential ($V$) the dynamical $w$ reduces to a constant value
\begin{equation}
w= \frac{\dot{\phi}^2 - \dot{\sigma}^2}{\dot{\phi}^2 - \dot{\sigma}^2}=1~.
\end{equation}
Thus the simplest case yields $p = \rho$, which corresponds to a radiation model and the evolution of both fields is determined by the forms above. Subsequently in all sub-cases of case 1, the equation of state possesses the same form.
\newline
\textbf{1a. (ii)} ($c_{i}\neq 0$ for all $i$)
\begin{eqnarray}
a(t) &= \frac{1}{2} \left (6c_{1}t^2 - 4c_{2}t + 8C_{1} \right ) ^{1/3}, \\
\sigma(t) &=  \frac{4c_{1}c_{5}}{ \sqrt{6(c_{4}^2 -c_{5}^2) }}\mbox{arctanh}{\left( \frac{3c_{1}t - c_{2}}{\sqrt{6(c_{4}^2 -c_{5}^2) }}\right)} + C_{2} ,\\
\phi(t) & = \frac{-4c_{4}c_{1}}{\sqrt{6(c_{4}^2 -c_{5}^2) }}\mbox{arctanh}{\left( \frac{3c_{1}t - c_{2}}{\sqrt{6(c_{4}^2 -c_{5}^2) }}\right)} + \frac{c_{3}-c_{4}C_{2}}{c_{5}} , \\
&\mbox{where} \quad C_{1} = \frac{c_{2}^2 - 6 (c_{4}^2 -c_{5}^2)}{12 c_{1}}, \quad |c_{4}|>|c_{5}|
\end{eqnarray}
\textbf{1b.} ($c_{i}\neq 0$ for all $i=2,..,5$, $c_{1}=0$)
\begin{eqnarray}
a(t) &= \frac{1}{2} \left ( C_{1}- 4c_{2}t \right ) ^{1/3}, \\
\sigma(t) &=  -\frac{2c_{5}}{ c_{2} }\ln \left ( C_{1}-c_{2}t\right)  + C_{2} ,\\
\phi(t) & =  \frac{2c_{4}}{ c_{2} }\ln \left ( C_{1}-c_{2}t\right)  + \frac{c_{3}-c_{4}C_{2}}{c_{5}}, \\
&\mbox{where} \quad c_{2} = \pm \sqrt{c_{4}^2-c_{5}^2}, \quad |c_{4}|>|c_{5}|
\end{eqnarray}
which yields two solutions relative to positive or negative roots of $c_{2}$. Note that both fields have opposite behavior for both roots while the behavior of arguments will be reversed if we take negative value of the root. The case $|c_{4}|<|c_{5}|$, gives imaginary form of the scale factor which was considered to be real. Hence we do not include it in our analysis.
\newline
\textbf{1c.} ($c_{i}=0$ for all $i=2,..,5$, $144c_{1}-3C_{1}^2=0$)
\begin{eqnarray}
a(t) &= \left( C_{1} t + C_{2}  \right )^{2/3}, \\
\sigma(t) &=  \mbox{const.} ,\\
\phi(t) & =  \mbox{const.} ,
\end{eqnarray}
which is a solution without dynamical fields. Note that for $C_{2}=0$, we obtain $a \propto t^{2/3}$ which is the  standard Friedman solution. Furthermore the constraint on $c_{1}$, ensures that all invariants are also satisfied. Since $I_{1}$ corresponds to time-translational symmetry with constant gauge, therefore $I_{1}=c_{1}$. The choice of this constant requires more understanding. In classical mechanics where a physical system is described by a Lagrangian that is the difference of a kinetic term and a potential term, the invariance of Lagrangian under time translation symmetry guarantee the conservation of energy (Hamiltonian). In this case we are not sure what this amounts to in our case as the given Lagrangian (\ref{lag1}) does not carry the same interpretation as in classical physics. On the other hand the Hamiltonian is always assumed to be zero in gravity and general relativity lacks to provide a precise understanding of the concept of energy of the universe. Without loss of generality we can redefine $C_{1}$ in terms of $c_{1}$, say $C_{1}= \pm \tilde{c}_{1},$ where $c_{1}=3\tilde{c}_{1}^2/144$. Hence we get
\begin{eqnarray}
a(t) &= \left( \pm \tilde{c}_{1} t + C_{2}  \right )^{2/3},
\end{eqnarray}
which gives us the same graph in either case ($\pm$), depending on the choice if $t$ is assumed to be positive or negative. The positive values of $C_{2},$ provides a translational shift to the graph on the left and on the right for negative values of $C_{2}$. It is pointed that although the form of potentials in above case is same as the form of potential in \cite{ad13} (Section 4), where they obtained an exponential form of the scale factor. It does not contradict our analysis. Because we can obtain the same form by taking the value $c_{1}=0$, i.e. when the ``Hamiltonian" (energy) is equated to zero. This case appears below for non-constant potential in a more general form.

\textbf{Case 2.} ($V(\phi,\sigma) = \mathcal{C} \neq 0$) \newline
In this case the scaling symmetry evaporates and we get same four Noether symmetries $X_{i}$, where $i=1,3,4,5$ and invariants. We first equate them to constants, $I_{i}=c_{i}$, where $i=1,3,4,5$. Now in this case the invariants $I_{4}$ and $I_{5}$, implies that
\begin{eqnarray}
\dot{\phi} = \frac{c_{4}}{a^3}, \quad \dot{\sigma} = -\frac{c_{5}}{a^3},
\end{eqnarray}
which reduces $I_{3}=c_{3}$, to a form that establishes a linear relationship between $\phi$ and $\sigma$, given by $c_{4}\sigma +c_{5}\phi+c_{3}=0$. The remaining invariant $I_{1}$, results into
\begin{equation}
3a^4 \dot{a}^2 - \mathcal{C}a^{6} - \frac{c_{4}^2 - c_{5}^2}{2} = c_{1} ,
\end{equation}
which can be simplified into
\begin{equation}
\dot{a} = \left ( \frac{\tilde{c} + \mathcal{C} a^6 }{3a^4}\right) ^{1/2}, \quad \tilde{c} = c_{1} + \frac{c_{4}^2-c_{5}^2}{2}.
\end{equation}
It can not be explicitly integrated. However, there arise a case in which if all gauge constants are zero $\tilde{c}=0$, then it results into following solution.
\newline
\textbf{2a.} ($c_{i}=0$ for all $i=1,3,4,5$)
\begin{eqnarray}
a(t) &= \mbox{exp}\left(\pm \sqrt{\frac{\mathcal{C}}{3}} (t-C_{1}) \right), \\
\sigma(t) &=  \mbox{const.} ,\\
\phi(t) & =  \mbox{const.} .
\end{eqnarray}
where it is clear that potential ($V=\mathcal{C}$) can not be negative and $C_{1}$ is an integration constant which produces a shift in the graph on left if it is positive and right if it is negative.  \newline
\textbf{Case 3.} ($V(\phi,\sigma) = (\phi\pm \sigma)^n$, $n\neq 0$) \newline
For this case, the Lagrangian (\ref{lag1}) admits three Noether symmetries (\ref{nsym1}). The action of the three dimensional group on the Lagrangian (\ref{lag1}), in the form of continuous transformations of constant parameters, is given below
\begin{eqnarray*}
X_{1} &= \partial_{t} \quad \longrightarrow \quad  L(t-\epsilon_{1}, a, \phi,\sigma) , \nonumber \\
X_{2} &= \partial_{\phi}\mp\partial_{\sigma} \quad \longrightarrow  \quad L(t, a, \phi \pm \epsilon_{2}, \sigma \mp \epsilon_{2}) ,\nonumber \\
X_{3} &= t\partial_{t} +\frac{a}{3}\partial_{a} \mp \frac{2}{n} \left ( \sigma \partial_{\phi}+\phi\partial_{\sigma}\right)  \longrightarrow  L(e^{\epsilon_{3}} t, e^{\epsilon_{3}/3}a, \tilde{\phi},\tilde{\sigma} ) , \nonumber \\
&\mbox{where, }  \tilde{\phi} = \cosh{\left(\frac{2\epsilon_{3}}{n}\right)} \phi \mp \sinh{\left(\frac{2\epsilon_{3}}{n}\right)} \sigma, \\
&\quad \quad \quad \tilde{\sigma} = \mp \sinh{\left(\frac{2\epsilon_{3}}{n}\right)} \phi + \cosh{\left(\frac{2\epsilon_{3}}{n}\right)} \sigma.
\end{eqnarray*}
Therefore scaling symmetry (which is a consequence of homogeneity) arise with the requirement that both fields $\phi$ and $\sigma$, undergo a hyperbolic rotation such that the difference of their squares $\phi^{2} -\sigma^{2}$, is preserved. It is emphasized that the dynamical invariants in both cases are crucial to investigate quintom models. Therefore we first bring them in a more suitable form to apply further analysis. It is interesting to see that $I_{3},$ in case 2, can be simplified into a form
\begin{eqnarray}
I_{3} = 2a^2 \dot{a} \mp \frac{2a^3}{n} ( \phi\dot{\sigma} - \sigma\dot{\phi} ) - tI_{1} ,
\end{eqnarray}
using the definition of $I_{1}$. Therefore we start with the assumption $c_{1}\neq 0$, so we obtain
\begin{eqnarray}
I_{3} = 2a^2 \dot{a}  \mp \frac{2a^3}{n} ( \phi\dot{\sigma} - \sigma\dot{\phi} )-  c_{1} t = c_{3} ,
\end{eqnarray}
where $c_{3}\neq 0$. In Cosmology, the equations are generally described in Hubble parameter, $H=\dot{a}/a$, in which case the above equation takes the form
\begin{eqnarray}
 2a^3 \left( H \mp \frac{1}{n} ( \phi\dot{\sigma} - \sigma\dot{\phi} )\right) -  c_{1}t  = c_{3} . \label{rinv1}
\end{eqnarray}
The other invariant $I_{2}$ is
\begin{eqnarray}
&I_{2} =  a^{3} \left( \dot{\phi} \pm \dot{\sigma} \right)=c_{2}. \label{rinv2}
\end{eqnarray}
Here we arrive at following subcases. \newline
\textbf{3a.} ($c_{2}=0$) \newline
Note that $c_{2}=0,$ implies that $\dot{\phi}= \mp \dot{\sigma}$, which can be integrated to get $\phi= \mp \sigma + C_{1}$, where $C_{1}$ is a constant of integration. Therefore equation (\ref{rinv1}) reduces to
\begin{eqnarray}
 2a^3 \left( H \mp \frac{C_{1}\dot{\sigma}}{n} \right) -  c_{1}t  = c_{3} .
\end{eqnarray}
This is a constraint equation on the equations of motions (\ref{eom}), as provided by the Noether gauge symmetry and help to reduce the dynamics. Note that it is a first order ordinary differential equation in two unknown functions and we require $C_{1}\neq 0$, as otherwise ($C_{1}=0$) the Lagrangian (\ref{lag}) becomes trivial. We solve equations of motions (\ref{eom}), along with above constraint equation and obtain following cases where evolution of all dynamical quantities is also given.
\newline
\textbf{3a. (i)} $V=(\phi+\sigma)^n$.\newline
\textbf{I-} ($n\neq 1,c_{1}=0, c_{3}=0$)
\begin{eqnarray}
a(t) &= C_{2} e^{\pm \sqrt{(C_{1}^n/3)}~ t}, \\
\sigma(t) & = \pm \left( \frac{C_{1}^{(n-2)}}{3}\right) ^{1/n} nt + C_{3} ,
\end{eqnarray}
\textbf{II-} ($n\neq 1,c_{1}=0, c_{3}\neq 0$)
\begin{eqnarray}
a(t) &= C_{2} e^{\pm \sqrt{(C_{1}^n/3)}~t}, \\
\sigma(t) & = \pm \frac{nc_{3}~e^{\mp \sqrt{3C_{1}^n}~ t}}{2C_{2}^{3}(3C_{1}^{n+2})^{1/2}} \pm \left(\frac{C_{1}^{n-2}}{3} \right)^{1/2} nt +C_{3} ,
\end{eqnarray}
\textbf{3a. (ii)} $V=(\phi-\sigma)^n$ \newline
\textbf{I-} ($n\neq 1,c_{1}=0, c_{3}=0$)
\begin{eqnarray}
a(t) &= C_{2} e^{\pm \sqrt{(C_{1}^n/3)}~ t}, \\
\sigma(t) & = \mp \left( \frac{C_{1}^{(n-2)}}{3}\right) ^{1/n} nt + C_{3} ,
\end{eqnarray}
\textbf{II-} ($n\neq 1,c_{1}=0, c_{3}\neq 0$)
\begin{eqnarray}
a(t) &= C_{2} e^{\pm \sqrt{(C_{1}^n/3)}~t}, \\
\sigma(t) & = \mp \frac{nc_{3}~e^{\mp \sqrt{3C_{1}^n}~ t}}{2C_{2}^{3}(3C_{1}^{n+2})^{1/2}} \mp \left(\frac{C_{1}^{n-2}}{3} \right)^{1/2} nt +C_{3} ,
\end{eqnarray}
\textbf{3b.} ($c_{2}\neq 0$) \newline
For this case we were unable to find any closed form of the solutions, however, invariants (\ref{rinv1}) and (\ref{rinv2}) provide us constraint equations to be satisfied by any quintom model. We have also verified that any possible solution to these equations yield a viable quintom model. This is mainly done by solving above equations (five of them) along with the condition that $w^{\prime}=0,$ from equation (\ref{dw}), it turned out that the entire system is inconsistent. Therefore all such quintom models require that $w^{\prime}\neq 0$.

\textbf{Case 4.} ($V(\phi,\sigma) = (\phi\mp \sigma)^n/(\phi\pm\sigma)^m$, $n\neq 0\neq m $) \newline
For the last case we first write invariants (\ref{inm}), in a proper form. For example we can use $I_{1}=c_{1}$, relative to
time translation to simplify $I_{2},$ in (\ref{inm}), where we get
\begin{eqnarray}
 2a^3 \left( \frac{\dot{a}}{a}\pm \frac{1}{n+m} ( \phi\dot{\sigma} - \sigma\dot{\phi} )\right) -  c_{1}t  = c_{3} . \label{rinv3}
\end{eqnarray}
Again in this case we were unable to find closed form of the solutions. However, equation (\ref{rinv3}), provide us a constraint equation to be satisfied by any viable quintom model.

\end{document}